\documentclass[aps,twocolumn,superscriptaddress,noshowkeys,amssymb,amsmath,amsfonts,longbibliography]{revtex4-1}

\usepackage{graphicx}
\usepackage{theorem}
\usepackage{dcolumn}
\usepackage{amssymb}
\usepackage{amsmath}
\usepackage{amsfonts}
\usepackage{bm}
\usepackage{tikz}
\usepackage{pdfpages}
\usepackage{verbatim}

\begin{document}

\title{Topological one-way fiber of second Chern number}

\author{Ling Lu}  \email{linglu@iphy.ac.cn}
\affiliation{Institute of Physics, Chinese Academy of Sciences/Beijing National Laboratory for Condensed Matter Physics, Beijing 100190, China}
\author{Zhong Wang}  \email{wangzhongemail@tsinghua.edu.cn}
\affiliation{Institute for Advanced Study, Tsinghua University, Beijing 100084, China}
\affiliation{Collaborative Innovation Center of Quantum Matter, Beijing 100871, China}

\begin{abstract}
Optical fiber is a ubiquitous and indispensable component in communications, sensing, biomedicine and many other lightwave technologies and applications.
Here we propose topological one-way fibers to remove two fundamental mechanisms that limit fiber performance: scattering and reflection.
We design three-dimensional~(3D) photonic crystal fibers, inside which photons propagate only in one direction, that are completely immune to Rayleigh and Mie scatterings and significantly suppress the nonlinear Brillouin and Raman scatterings.
A one-way fiber is also free from Fresnel reflection, naturally eliminating the needs for fiber isolators.
Our finding is enabled by the recently discovered Weyl points in a double-gyroid~(DG) photonic crystal.
By annihilating two Weyl points by supercell modulation in a magnetic DG, we obtain the photonic analogue of the 3D quantum Hall phase with a non-zero first Chern number~($C_1$).
When the modulation becomes helixes, one-way fiber modes develop along the winding axis, with the number of modes determined by the spatial frequency of the helix.
These single-polarization single-mode and multi-mode one-way fibers, having nearly identical group and phase velocities, are topologically-protected by the second Chern number~($C_2$) in the 4D parameter space of the 3D wavevectors plus the winding angle of the helixes. This work suggests a unique way to utilize higher-dimensional topological physics without resorting to artificial dimensions.
\end{abstract}
\maketitle

Topological photonics~\cite{lu2014topological,lu2016topological} started with the realization of one-way edge waveguides~\cite{haldane2008possible,wang2009observation,fang2012realizing,rechtsman2013photonic,hafezi2013imaging,chen2014experimental,wu2015scheme,cheng2016robust,fu2010robust,poo2011experimental,he2016photonic,iadecola2016non} as the analog of chiral edge states in quantum Hall effect, where the number and direction of 1D edge modes~\cite{skirlo2015experimental} are determined by the 2D bulk topological invariant: the first Chern number~($C_1$). These Berry monopoles has now been realized in 3D as Weyl points~\cite{lu2015experimental}, opening doors to 3D topological phases for photons~\cite{lu2016symmetry}.
Here we show that, by annihilating a single pair of Weyl points with helix modulations, light can be guided unidirectionally in the core of 3D photonic crystal fibers where the number and direction of one-way modes equals the magnitude and sign of the second Chern number~($C_2$) --- the topological invariant of complex vector bundles on 4D manifolds.
In addition, we provide a definitive way to obtain arbitrary mode number~(from -$\infty$ to +$\infty$) in the one-way fibers by varying the helix frequencies. Furthermore, all the modal dispersions have almost identical group and phase velocities, superior for multimode operations. The above aspects indicate that topological one-way fibers are fundamentally distinct and more advantageous than the topological one-way edge waveguides in 2D.

\begin{figure}[]
\includegraphics[width=0.45\textwidth]{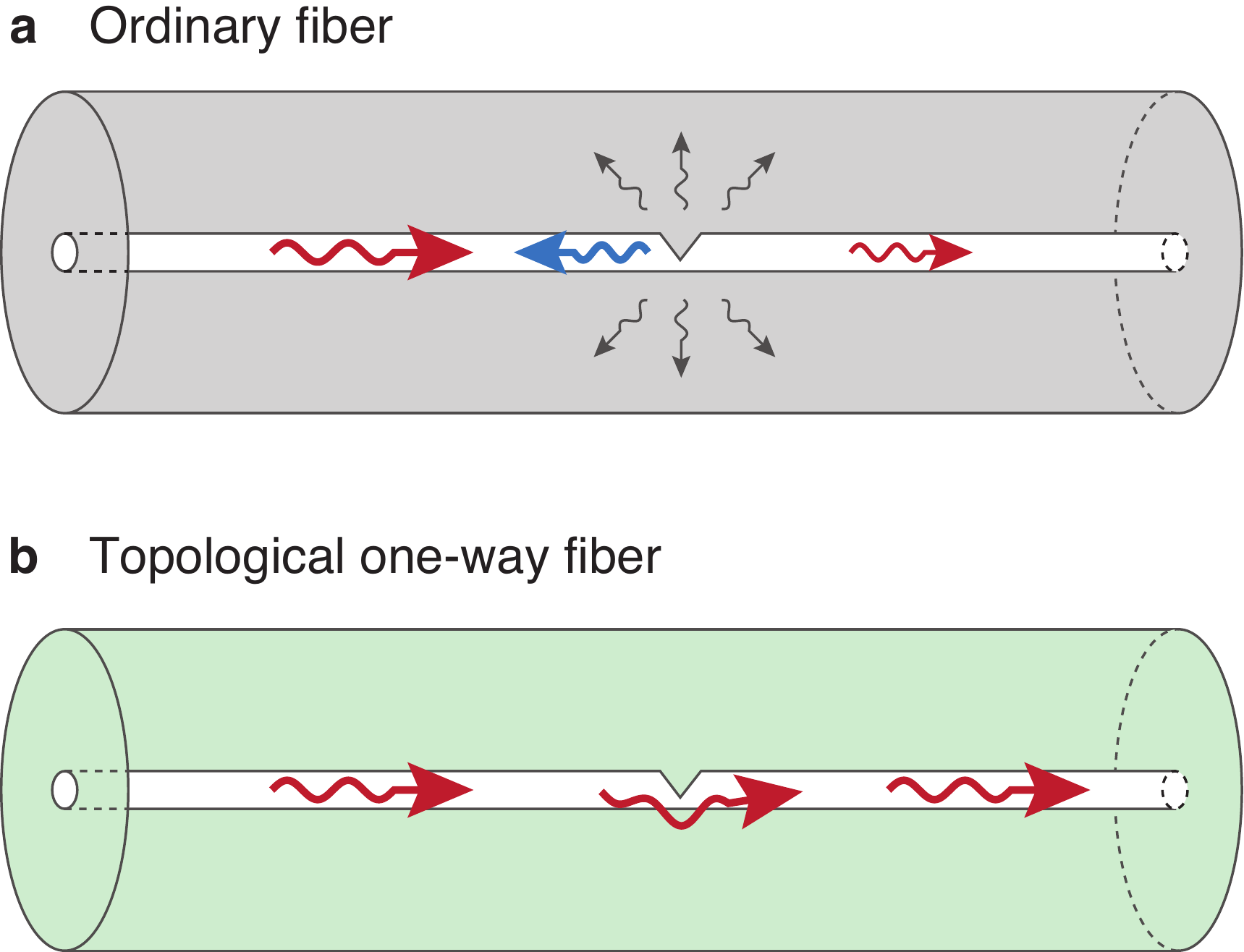}
\caption{Ordinary fiber versus topological one-way fiber.
a) The optical mode in an ordinary fiber is confined in the core, where both forward and backward modes exist. The imperfection induces scattering losses into back and radiation channels.
(b) The optical mode in a topological one-way fiber is confined by the 3D topological bandgap which spatially separates the forward and backward modes. The one-way mode in the core is immune to scattering losses of any kinds.}
\label{fig1}
\end{figure}

Topological one-way fiber represents a conceptual leap in fiber designs~
\cite{kao1966dielectric,temelkuran2002wavelength,russell2003photonic,dudley2006supercontinuum} (Fig. \ref{fig1}), potentially advancing many aspects of fiber technologies.
Firstly, Rayleigh scattering sets the ultimate lower limit for fiber loss.
In the absence of Rayleigh scattering, topological one-way fiber only suffer from material absorption and could enable ultra-low loss fibers.
Secondly, Mie scattering, occurring at irregular interfaces and miro-bendings, is mostly suppressed in topological fibers -- another way of lowering fiber loss.
Thirdly, stimulated Brillouin scattering~(SBS) back-couples the forward signal and usually restricts the fiber transmitting power to a few $mW$s. This SBS threshold sets the upper bound on the maximum injection power in fiber communication systems, limiting the signal-to-noise ratio and the transmission distance without amplification. SBS threshold can now be lifted in one-way fibers.
Furthermroe, half of the Raman scattering~(backward portion) is also inhibited. 
Fourthly, Fresnel reflections at splices and connectors disappear in one-way fibers, so that fiber connectors no longer need to be specially designed to minimize back reflections. More importantly, one-way fibers completely remove the necessity of fiber isolators -- a bulky but essential component in any fiber systems.
Lastly, optical force inside a topological one-way fiber is conservative~\cite{wang2011response}, i.e., there is no propelling force for objects inside the hollow-core due to the absence of back-scattered photons. This could be used for optical trapping in fibers.

\begin{figure}[]
\includegraphics[width=0.5\textwidth]{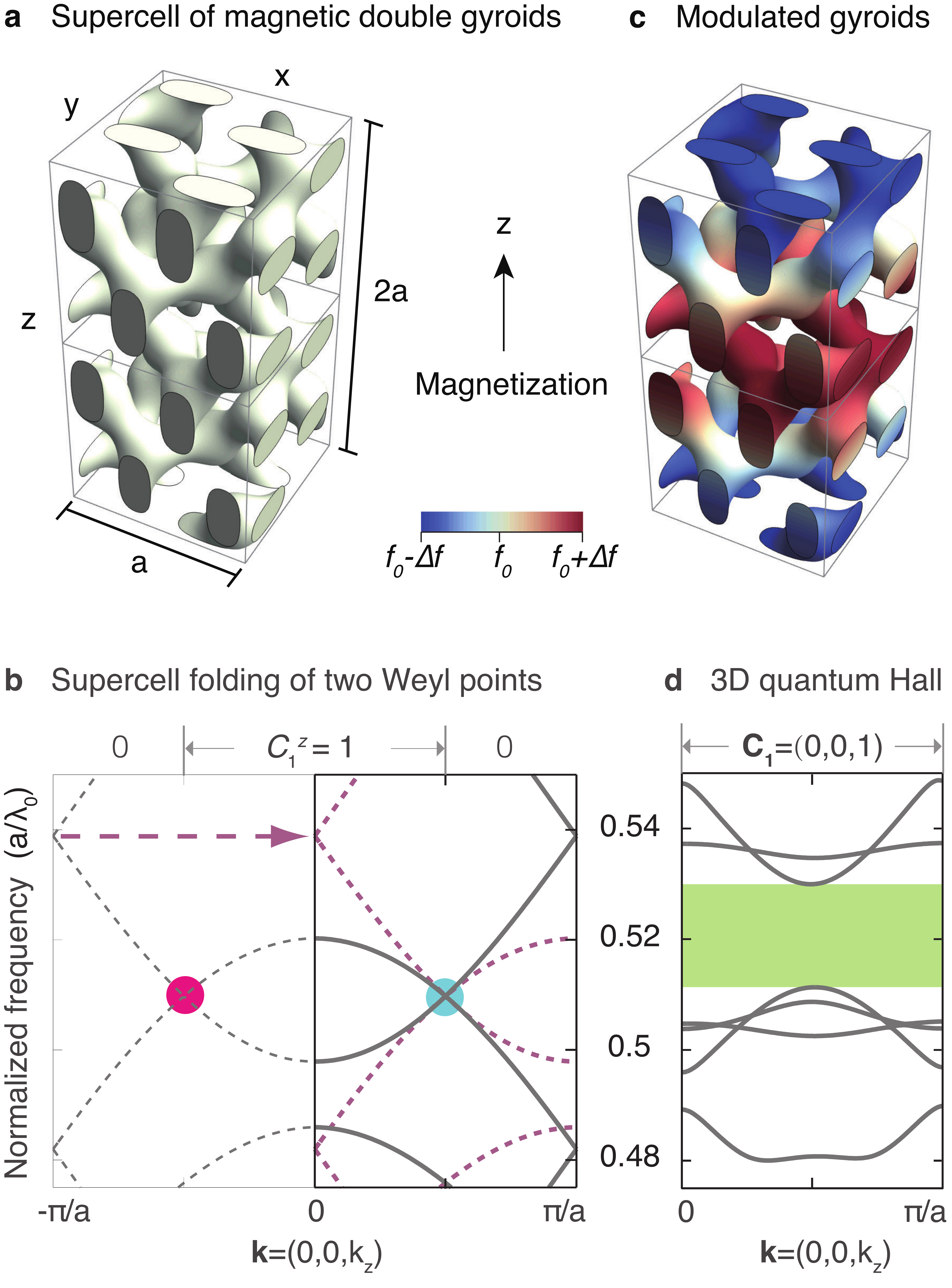}
\caption{ (a) Two cubic unit-cells of the DG photonic crystal magnetized along $z$. (b) The band structure of a cubic cell shows two Weyl points, which fold into one 3D Dirac point in the Brillouin zone of the supercell. (c) The DG photonic crystal whose volume fraction~(colored) is periodically modulated along $z$. (d) The band structure of the photonic analogue of 3D Chern insulator whose topological gap frequencies are highlighted in green.}
\label{fig2}
\end{figure}

Our starting point is a photonic crystal containing two Weyl points~\cite{lu2013weyl}, which were found in the double gyroids~(DG) made of magnetic materials~(either gyroelectric or gyromagnetic).
The DG is a minimal surface that can be approximated by the iso-surface of a single triply-periodic functions:
$
f(x,y,z)=\sin(2\pi x/a)\sin(2\pi y/a)\cos(4\pi z/a) + \sin(2\pi y/a)\sin(2\pi z/a)\cos(4\pi x/a) + \sin(2\pi z/a)\sin(2\pi x/a)\cos(4\pi y/a)
$.
 This definition yields almost identical geometry and band structure to those of the DG defined in Ref. \cite{lu2013weyl} by two trigonometric functions~(one for each gyroid).
Shown in Fig. \ref{fig2}a are two cubic cells of the DG, where $f(x,y,z)>f_0=0.4$ is filled with gyroelectric material of dielectric constant $\epsilon=
\begin{pmatrix}
17 & -6i & 0\\
+6i & 17 & 0\\
0 & 0 & 16
\end{pmatrix}
$ and unity magnetic permeability. The rest of the volume is air.
In this structure, there exists only two Weyl points~(a single ``Weyl dipole'') separated by \emph{about} half of the Brillouin zone along $z$ direction, as plotted in Fig. \ref{fig2}b.
This means that a real-space modulation of the crystal, in $z$ with a period of $2a$, can couple the two Weyl points and open up a complete 3D bandgap.
The fact that a bandgap does not close under small perturbations ensures the robustness of this approach: certain mismatch between the Weyl-point separation and the wavevector of the modulation can be tolerated.

\emph{3D quantum Hall phase.---}We create a double-cell periodic modulation along $z$ to annihilate the Weyl points and obtained the 3D quantum Hall phase~\cite{halperin1987possible,koshino2001hofstadter,bernevig2007theory}, also known as the 3D Chern insulator. (So far, experimental realization of 3D quantum Hall phase is only limited to quasi-2D systems~\cite{stormer1986quantization}.)
This modulation can be implemented in various system parameters such as volume fraction, refractive index, magnetization or structural distortion. In this work, we modulated the volume fraction of the DG by modifying the DG equations as follows:
$f(x,y,z) > f_0 + \Delta f\cos(\pi z/a)$, in which $\Delta f=0.07$.
The modulated DG is shown in Fig. \ref{fig2}c and its band structure plotted in Fig. \ref{fig2}d.

A 3D quantum Hall phase is characterized by three first Chern numbers $\mathbf{C_1}=(C_1^x, C_1^y, C_1^z)$ defined on the $\hat{x}$, $\hat{y}$ and $\hat{z}$ momentum planes.
For example, $C_1^z$ is defined as
\begin{eqnarray}
C_{1}^{z}\equiv C_{1}^{z}(k_z)=\frac{1}{2\pi}\int d^2k \rm{Tr} [ \mathcal{F}_{xy} ].
\label{eq::C1}
\end{eqnarray}
Because the bulk spectrum is gapped, $C_1^z$ cannot change as a function of $k_z$.
When there are $N$ bulk bands below the bandgap, $\mathcal{F}_{xy}$ is an $N\times N$ matrix, whose elements are $\mathcal{F}^{\alpha\beta}_{xy}=\partial_x \mathcal{A}^{\alpha\beta}_y-\partial_y
\mathcal{A}^{\alpha\beta}_x+i\left[\mathcal{A}_x,\mathcal{A}_y\right]^{\alpha\beta}$, in which $\alpha, \beta = 1, 2, \cdots ,N$.
The Berry connection $\mathcal{A}_i^{\alpha\beta}(\mathbf{k}) = -i\left\langle \psi^\alpha(\mathbf{k})\right| \frac{\partial}{\partial k_i } \left|\psi^\beta(\mathbf{k})\right\rangle$, where $|\psi^{\alpha(\beta)} \rangle$ are the Bloch eigenfunctions.
Note that the trace of the commutator $\rm{Tr}\left[\mathcal{A}_x,\mathcal{A}_y\right]$ always vanishes for the first Chern class.

The topological invariants of our modulated DG is $\mathbf{C_1}$=(0,0,1).
This can be understood from the original Weyl photonic crystal whose first Chern number is one for half of its Brillouin zone, as illustrated in Fig. \ref{fig2}b. By folding the Brillouin zone to half of its original size, the Chern numbers in different regions add up (Fig. \ref{fig2}b).

3D quantum Hall phase is a weak topological phase whose weak topological invariants are defined in a lower dimension as compared to a strong topological phase with strong topological invariant.
It is theoretically known that a lattice dislocation in a weak topological phase creates a 1D topological defect mode~\cite{ran2009one}.
Unfortunately, in our case, a dislocation induces significant lattice distortion that generates many additional non-topological modes in the bandgap.

Fortunately, we propose and demonstrate below that, for a 3D quantum Hall phase constructed from Weyl crystals, a new approach is available: a smooth helical modulation generates a one-way mode at the core of the helix. The advantage here, compared with the dislocation approach, is the intactness of the lattice that prevents the generation of non-topological modes in the bandgap.
Before presenting rigorous calculations, we outline a physical interpretation as follows (see Supplemental Material for details).
A supercell modulation couples two Weyl points of opposite chiralities, forming a gapped 3D Dirac point with a mass term that is complex-valued. (A 3D Dirac point consists of two Weyl points of opposite chiralities.)
Then a helical modulation amounts to a nonzero winding number of the phase of the Dirac mass around the helical axis, and it was indicated in previous theoretical models that such a topological perturbation can generate topological defect modes in both 2D \cite{jackiw1981zero,hou2007electron} and 3D systems \cite{callan1985anomalies,bi2015unidirectional,schuster2016dissipationless}.

\begin{figure*}[t!]
\includegraphics[width=\textwidth]{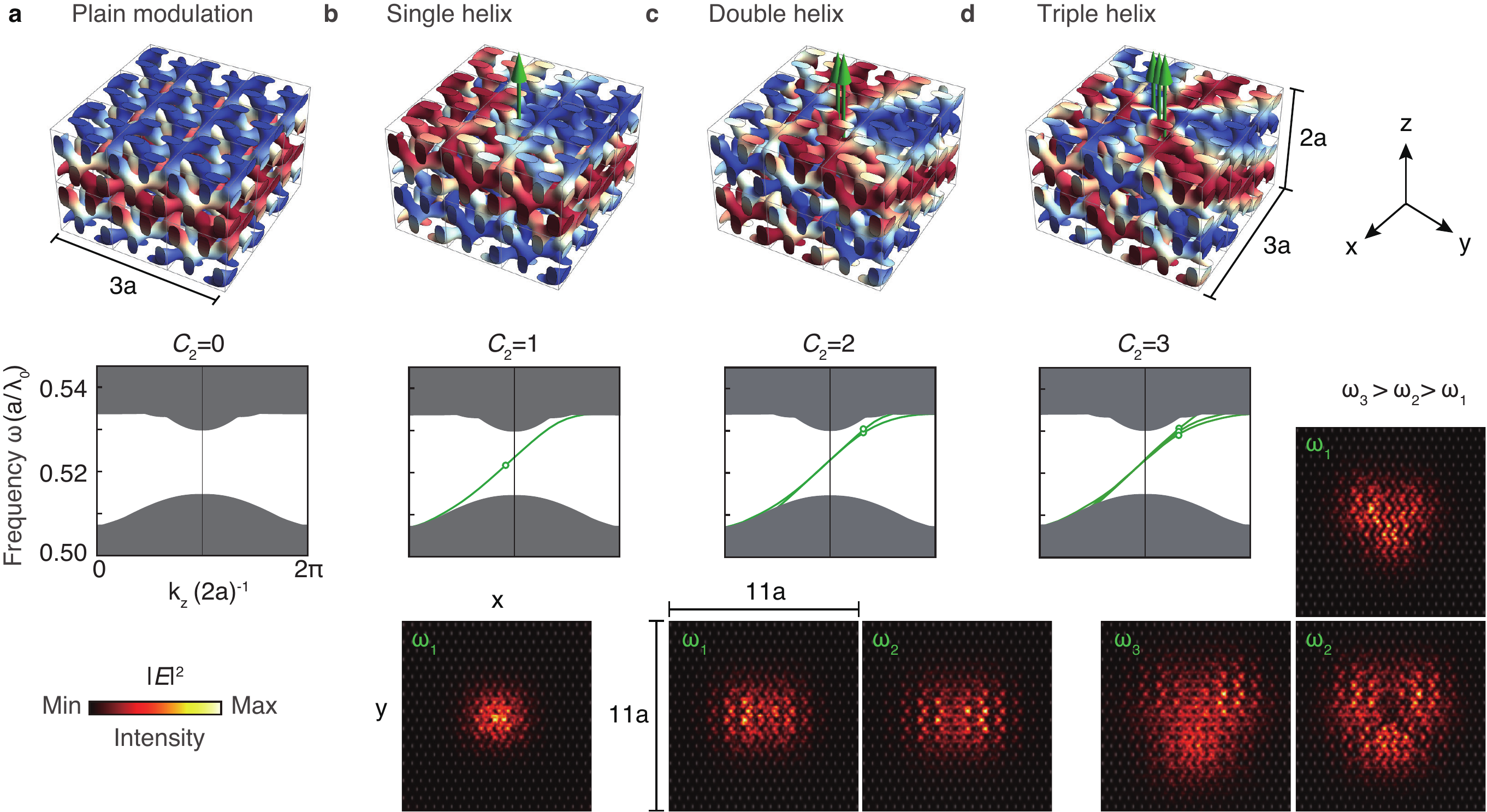}
\caption{
Single and multi-mode fibers in helically-modulated magnetic DGs.
a) The DG structure of a plain modulation~($w=0$), shown in a $3\times3\times2$ cubic cells, and its projected band structure.
b) The single helix DG structure~($w=1$), whose helix center supports one one-way fiber mode.
c) The double helix DG structure~($w=2$), whose helix center supports two one-way fiber modes.
d) The triple helix DG structure~($w=3$), whose helix center supports three one-way fiber modes.
The fiber dispersions and mode profiles are plotted below the plots of their structures.
Each dispersion plot is labeled by the second Chern number~($C_2$, to be defined below) of the system, which equals $w$ in these examples.
The mode profiles of $C_2=2$ and $C_2=3$ look more extended than that of $C_2=1$.
The reason is that we plotted the modal profiles for $C_2=2,3$ at frequencies close to the bulk bands, where the modal confinement is weaker than that of the modes closer to the central frequencies of the bandgap. 
We did not plot the modes at the gap center~(as for $C_2=1$), 
because all modes there have almost the same wavevector and frequency, which enhances the modal coupling and results in difficulty in resolving their intrinsic field patterns.
}
\label{fig3}
\end{figure*}

\emph{One-way fiber modes.---}Now comes the crucial step in our design of topological one-way fibers. Instead of the plain modulation (fig. \ref{fig2}c), we create a helical modulation by filling the volume defined by the equation:
\begin{eqnarray}
f(x,y,z) > f_0 + \Delta f\cos(\pi z/a + w\theta)
\label{eq::chiralDG}
\end{eqnarray}.
The modulation now winds as a function of the angle $\theta$~[$\arctan(y/x)$] in the $x-y$ plane, whose spatial frequency is controlled by the signed integer $w$. 
The sign and magnitude of $w$ determines the direction and number of the one-way modes on the winding axes.
This is illustrated in the upper panels of Fig. \ref{fig3} for $w=+1,+2,+3$, corresponding to single, double and triple helix one-way fibers.

The band structures of the one-way fibers are shown in  middle panels of Fig. \ref{fig3}. They were calculated using MIT Photonic Bands on an $11\times 11\times 2$ cubic supercell.
The spectra  exhibit one-way modes within the bulk band gap. The profiles of the topological fiber modes are localized around the helix cores (Fig. \ref{fig3} lower panel).

Shown in the middle panel of Fig. \ref{fig3}, all one-way-fiber dispersions~(green lines) have very similar phase and group velocities. In the multimode cases, their dispersions almost overlap on top of each other. This is due to the fact that these defect modes originated from the same Weyl bulk bands, so they all share the same Brillouin-zone location and group velocities as that of the original Weyl cones. This behavior is different from that of the one-way edge waveguides in 2D~\cite{skirlo2015experimental}, where the edge modal dispersions have different phase or group velocities. This can be attributed to the fact that the edge environment is distinct from the environment of the 2D bulk lattice, while, here, there is no sharp interfaces in the 3D one-way fibers.
This unique feature, of multiple fiber modes having almost identical dispersions, ensures multimode signals propagate at the same speed without intermodal distortion.

Hollow-core one-way fibers can be made by removing the materials on the helix axis, in which the mode confinement in the air core is ensured by the 3D photonic bandgap. The extra air volume in the hollow core may generate trivial fiber modes, which can be tuned away. In contrast, the existence of the non-trivial one-way mode is protected by the topological invariant of the system.

\emph{Second Chern number.---}It is natural to ask for a topological invariant for the one-way fibers.
With the simplest helix modulation of the form of Eq. \ref{eq::chiralDG}, it is intuitive to guess that $w$ is the topological invariant, since the number and direction of the one-way modes match the magnitude and sign of $w$.
However, this observation does not work if we consider the modulation of the general form as $f(x,y,z) > 0.4 + \sum_w h_w\cos(\pi z/a + w\theta)$, where $h_w$ are real-valued constants.

For a \emph{lattice} dislocation in a 3D Chern insulator~(the 3D QHE), it is known\cite{teo2010topological} that the number of chiral modes is given by ${\bf C}_1\cdot{\bf b}$, where the dimensionless Burgers vector(${\bf b}$) represents the magnitude and direction of the lattice distortion.
However, this approach can not be applied to our system due to the lack of unique ``Burgers vector'' other than in the simplest case~(as of Eq. \ref{eq::chiralDG}).

We show that the desired topological invariant of our one-way fibers are the second Chern number~($C_2$), the strong topological invariant in our system.
Note that, far away from the axis of the helix, the Bloch Hamiltonian smoothly varies with $\theta$, thus is a smooth function of the four variables $(k_x,k_y,k_z,\theta)$. Since $(k_x,k_y,k_z, \theta)$ span a four-dimensional parameter space with periodic boundary conditions~(a 4D torus), the second Chern number\cite{Qi2008,teo2010topological} can be defined:  
\begin{eqnarray}
C_2=\frac1{4\pi^2}\int d^3k d\theta
\rm{Tr}\left[\mathcal{F}_{xy} \mathcal{F}_{z\theta}+\mathcal{F}_{yz} \mathcal{F}_{x\theta}+\mathcal{F}_{zx} \mathcal{F}_{y\theta} \right].
\label{eq::C2} \end{eqnarray}
Similar to the definitions in Eq. \ref{eq::C1}, $\mathcal{F}^{\alpha\beta}_{ij}=\partial_i \mathcal{A}^{\alpha\beta}_j-\partial_j
\mathcal{A}^{\alpha\beta}_i+i\left[\mathcal{A}_i,\mathcal{A}_j\right]^{\alpha\beta}$, in which $\alpha,\beta$ are the band indices.
The non-Abelian Berry potential $\mathcal{A}_i^{\alpha\beta}(\mathbf{k},\theta) = -i\left\langle \psi^\alpha(\mathbf{k},\theta)\right| \frac{\partial}{\partial k_i } \left|\psi^\beta(\mathbf{k},\theta)\right\rangle$, where $|\psi^{\alpha(\beta)} \rangle$ are the eigenfunctions and $k_i$ runs through $k_x,k_y,k_z,\theta$. It is notable that this definition of $\mathbf{C_2}$ involves three variables ($k_{x,y,z}$) in the reciprocal space and one variable ($\theta$) in the real space, in contrary to the four momentum variables in the 4D quantum Hall effect\cite{zhang2001four,Qi2008,kraus2013four,price2015four}.
Consequently, the Berry curvature $\mathcal{F}_{i\theta}$ is even while $\mathcal{F}_{ij}$ is odd under time-reversal, where $i$ or $j$ represents one of $x$,$y$ and $z$.
Although in 4D QHE, $\mathbf{C_2}$ can be non-zero without breaking time-reversal symmetry, non-zero $\mathbf{C_2}$ requires time-reversal breaking in our system.

In the Supplemental Material, we carried out the explicit calculations of $C_2$, which is consistent with our numerical findings in Fig. \ref{fig3}. The topological protection by the second Chern number indicates that the physical origin of the one-way fiber modes is fundamentally different from that of the edge modes of the photonic analog of 2D quantum Hall effect\cite{haldane2008possible,wang2009observation}, whose topology is captured by the first Chern number. We note that although, in our system, both the weak indices~($\mathbf{C_1}$) and the strong index~($C_2$) are non-zero, it is possible to construct a one-way fiber design with zero $\mathbf{C_1}$ and non-zero $C_2$. For example, 
when the separation between the two Weyl points shrinks zero (forming a 3D Dirac point), one can apply only angular ($\theta$) modulations to obtain a one-way fiber of non-zero  $\mathbf{C_2}$ but zero $\mathbf{C_1}$s.

\emph{Experimental feasibility.---}
Various existing technologies can be adopted to realize these magnetic fibers in experiments across various frequences. At microwave frequencies, the sample can be fabricated by the same drilling and stacking approach as demonstrated in Ref. \cite{lu2015experimental} using gyromagnetic materials~\cite{wang2009observation,skirlo2015experimental}.
Similar methodology can be used at terahertz wavelengths.
Towards optical frequencies, gyroelectrical materials are the choices.
For example, paramagnetic Terbium-doped magnetic fibers have already been demonstrated with high Verdet constants~\cite{sun2010compact,schmidt2011complex}.
Ferrimagnetic materials, having a stronger magneto-optical effect than paramagnetic ones, usually suffer higher optical losses. Fortunately, they are continually being improved~\cite{onbasli2016optical} and enhanced~\cite{luo2016magneto}. 
A DG fiber can either be made by drawing a 3D-printed preform or potentially by self-assembly~\cite{hur2011three,khoo2013blue} during the drawing process.
3D direct writing~\cite{turner2013miniature} and interference lithography~\cite{ullal2003triply} can also be adopted.
Finally, the chiral modulation can be created by spinning the fiber during drawing, as demonstrated in the chiral fibers~\cite{kopp2011chiral,wong2012excitation}.

\emph{Outlook.---}The proposal of one-way fibers enriches the prospects of device applications for the Weyl materials, topological photonics and topological physics in general. It also brings a new playground on the realization of higher dimensional topological phases.
The same phenomena can be realized in other Weyl systems~\cite{wan2011,xu2015discovery,lv2015experimental,wang2016topological,chen2016photonic,xiao2015synthetic,rocklin2015mechanical,gao2015plasmon,li2016weyl} with time-reversal symmetry breaking.
More importantly, such topological fibers can inspire new directions, design principles and applications for fiber technology.

\emph{Acknowledgements.---}We thank the discussion with Jian Wang, Wei Ding and Changyuan Yu on fiber technologies and with Hannah M. Price and Chen Fang on 3D QHE.
Z. W. was supported by NSFC under Grant No. 11304175.
L.L was supported in part by the Ministry of Science and Technology of China under Grant No. 2016YFA0302400.
L. L. was supported in part by the National Thousand-Young-Talents Program of China.
Part of the numerical calculations of this work was performed using MIT Photonic Bands on En-Real.com.

\clearpage


\section*{Supplementary material}

# Supplemental Material for "Topological one-way fiber of second Chern number"

Ling Lu[1, *] and Zhong Wang[2, 3, †]

[1]*Institute of Physics, Chinese Academy of Sciences/Beijing National Laboratory for Condensed Matter Physics, Beijing 100190, China*
[2]*Institute for Advanced Study, Tsinghua University, Beijing 100084, China*
[3]*Collaborative Innovation Center of Quantum Matter, Beijing 100871, China*

In the main text we have presented our design of the one-way fiber and the results of band-structure calculations. To gain a simple analytical understanding of the one-way modes, we outline below an effective Hamiltonian description, using the low-energy Dirac Hamiltonian. The picture can be summarized as follows. In terms of the effective Dirac Hamiltonian, the modulation corresponds to the presence of a Dirac mass. The helix modulation introduces a topologically nontrivial configuration of the Dirac mass (nonzero winding number of the phase of Dirac mass), which creates topologically one-way modes.

### Effective 3D Dirac Hamiltonian

In the absence of modulations, the double gyroid crystal host two Weyl points with opposite chirality, either of which is described by a $2 \times 2$ effective Weyl Hamiltonian. We can combine them as a $4 \times 4$ block-diagonal Dirac Hamiltonan:

$$H_{\rm D} = -iv(\sigma_x \partial_x + \sigma_y \partial_y + \sigma_z \partial_z)\tau_z, \qquad (1)$$

where $\sigma_i, \tau_i$ ($i = x, y, z$) are Pauli matrices, $v$ is the group velocity (for simplicity, we take isotropic group velocities with $v > 0$). Here $\tau_z = 1$ and $\tau_z = -1$ is associated with the two Weyl points, respectively. As we have seen in the main text, a modulation with periodicity $2a$ in the $z$ direction gaps out the Weyl points. In the effective Dirac Hamiltonian description, the frequency gap is due to the Dirac mass terms. There are only two possible Dirac mass terms that can generate a frequency gap: $m_1 \tau_x$ and $m_2 \tau_y$, both of which anti-commute with the $\sigma_i \tau_z$ terms in $H_{\rm D}$. It is thus expected that the modulation amounts to the presence of these Dirac mass terms in the low energy effective Hamiltonian. In general, both $m_1$ and $m_2$ can be nonzero, and the mass term can be written as $m_1 \tau_x + m_2 \tau_y = m\tau_+ + m^* \tau_-$, with $\tau_\pm \equiv (\tau_x \pm i\tau_y)/2$ and $m \equiv m_1 - im_2$. The full effective Hamiltonian, with the effect of modulation included as the Dirac mass, can be written as

$$H_{\rm eff} = -iv(\sigma_x \partial_x + \sigma_y \partial_y + \sigma_z \partial_z)\tau_z + m\tau_+ + m^* \tau_-, \qquad (2)$$

from which we can readily see that a frequency gap $2|m|$ is generated by the Dirac mass terms.

The phase factor of the complex-valued Dirac mass $m$ is crucial and deserves more discussions. Suppose that the modulation can be modelled in the effective Hamiltonian by a perturbation $V(\mathbf{r}) = V_\mathbf{Q} \exp(i\mathbf{Q}\cdot\mathbf{r}) + V_\mathbf{Q}^\dagger \exp(-i\mathbf{Q}\cdot\mathbf{r}) + \cdots$, where $\mathbf{Q} = (0, 0, \pi/a)$ is the wave vector of the modulation that couples the two Weyl points. We can readily see that $\exp(\pm i\mathbf{Q}\cdot\mathbf{r}) \rightarrow \tau_\pm$ is valid near the Weyl points, thus we have $m = V_\mathbf{Q}$, in other words, the complex-valued Dirac mass is simply the $\mathbf{Q}$-component of the perturbation.

Now we have the following important observation. If we displace the modulation by a distance $\mathbf{d}$, then the perturbation becomes $V(\mathbf{r} + \mathbf{d})$, which can be expand as $V(\mathbf{r} + \mathbf{d}) = V_\mathbf{Q} \exp[i\mathbf{Q}\cdot(\mathbf{r}+\mathbf{d})] + V_\mathbf{Q}^\dagger \exp[-i\mathbf{Q}\cdot(\mathbf{r}+\mathbf{d})] + \cdots$, thus we can see that the displacement causes $V_\mathbf{Q} \rightarrow V_\mathbf{Q} \exp(i\mathbf{Q}\cdot\mathbf{d})$, or equivalently, $m \rightarrow m\exp(i\mathbf{Q}\cdot\mathbf{d})$. For the helix-shape modulation along the axis $r = 0$, as described in the main text, the displacement $\mathbf{d}$ is a function of $\theta$ such that $\mathbf{Q}\cdot\mathbf{d} = w\theta$[3], therefore, we have a nonzero winding of the phase of Dirac mass around the axis, namely, $m(\theta) = m_0 \exp(iw\theta)$, in which $m_0 \equiv m(\theta = 0)$. The overall phase of $m_0$ can be changed by rotating the coordinate systems around the $r = 0$ axis, thus we are free to take $m_0$ to be real-valued and positive.

### Analytic solutions of the one-way modes

For the effective Dirac Hamiltonian with a nonzero winding of the phase of Dirac mass (with winding number $w$), which is a consequence of the helix perturbation, we shall show that there exist $|w|$ topological one-way modes. We can rewrite Eq.(2) in the cylindrical coordinates as

$$H_{\text{eff}} = \begin{bmatrix} vk_z, & -ive^{-i\theta}(\frac{\partial}{\partial r} - \frac{i}{r}\frac{\partial}{\partial \theta}), & m_0 e^{iw\theta}, & 0 \\ -ive^{i\theta}(\frac{\partial}{\partial r} + \frac{i}{r}\frac{\partial}{\partial \theta}), & -vk_z, & 0, & m_0 e^{iw\theta} \\ m_0 e^{-iw\theta}, & 0, & -vk_z, & ive^{-i\theta}(\frac{\partial}{\partial r} - \frac{i}{r}\frac{\partial}{\partial \theta}), \\ 0, & m_0 e^{-iw\theta}, & ive^{i\theta}(\frac{\partial}{\partial r} + \frac{i}{r}\frac{\partial}{\partial \theta}) & vk_z \end{bmatrix}. \tag{3}$$

where we have taken advantage of the translational symmetry in the $z$ direction by replacing $-i\partial_z$ by $k_z$. For notational simplicity, we shall keep implicit the common factor $\exp(ik_z z)$ in the eigenfunction. For reason that will become clear shortly, we look for eigenfunctions of the form of $\psi = [\psi_1, 0, 0, \psi_4]^T$. The eigenvalues are $E = vk_z$, and the eigenfunctions satisfy

$$-ive^{i\theta}(\frac{\partial}{\partial r} + \frac{i}{r}\frac{\partial}{\partial \theta})\psi_1 + m_0 e^{iw\theta}\psi_4 = 0,$$
$$m_0 e^{-iw\theta}\psi_1 + ive^{-i\theta}(\frac{\partial}{\partial r} - \frac{i}{r}\frac{\partial}{\partial \theta})\psi_4 = 0. \tag{4}$$

It is not difficult to observe that the second equation is equivalent to the first one if we take $\psi_4 = \pm\psi_1^*$. Let us focus on the $\psi_4 = \psi_1^*$ case first. With this condition, the above two equations are reduced to a single equation

$$-ive^{i\theta}(\frac{\partial}{\partial r} + \frac{i}{r}\frac{\partial}{\partial \theta})\psi_1 + m_0 e^{iw\theta}\psi_1^* = 0 \tag{5}$$

For the $w = 1$ case, the common $\exp(i\theta)$ factors can be eliminated, thus the equation becomes especially simple, and the solution can be found analytically as

$$|\psi_{w=1}\rangle = \begin{pmatrix} e^{i\pi/4} \\ 0 \\ 0 \\ e^{-i\pi/4} \end{pmatrix} e^{-\frac{m_0}{v}r}. \tag{6}$$

For an arbitrary integer $w \geq 1$, by analysis analogous to Ref.[1], we can show that there exist $w$ localized modes. In fact, we can take the following ansatz for Eq.(5):

$$\psi_1^{(l)} = e^{i\pi/4}[u_l e^{il\theta} + v_l e^{i(w-1-l)\theta}], \tag{7}$$

with an integer parameter $l$, whose acceptable values are to be determined. We first notice that when $l = (w-1)/2$, $e^{il\theta}$ and $e^{i(w-1-l)\theta}$ are actually equal, and the $v_l$ term is redundant. Let us first focus on the cases $l \neq (w-1)/2$. The special case $l = (w-1)/2$, with $e^{il\theta} = e^{i(w-1-l)\theta}$, will be discussed separately later.

According to Eq.(5), the coefficient functions $u_l, v_l$ have to satisfy

$$v(\frac{\partial}{\partial r} - \frac{l}{r})u_l + m_0 v_l = 0$$
$$v(\frac{\partial}{\partial r} - \frac{w-1-l}{r})v_l + m_0 u_l = 0. \tag{8}$$

The asymptotic behaviors of $u_l, v_l$ in the $r \to 0$ limit can be found as (I) $u_l \sim r^l, v_l \sim r^{l+1}$ or (II) $u_l \sim r^{w-l}, v_l \sim r^{w-l-1}$. On the other hand, the asymptotic behaviors in the $r \to \infty$ limit are (a) $u_l \to \exp(-\frac{m_0}{v}r), v_l \to \exp(-\frac{m_0}{v}r)$ or (b) $u_l \to \exp(\frac{m_0}{v}r), v_l \to \exp(\frac{m_0}{v}r)$, only the first of which is normalizable in the $r \to \infty$ regime. A normalizable solution must have behavior (a) in the $r \to \infty$ limit, which is generally a superposition of (I) and (II) in the $r \to 0$ regime. Therefore, the normalizability of the solution in the $r \to 0$ limit requires that both (I) and (II) are normalizable, which leads to the constraint

$$0 \leq l \leq w - 1. \tag{9}$$

Thus we have proved that, leaving out the special case $l \neq (w-1)/2$ undetermined, there exists one normalizable solution for every integer $l = 0, 1, 2, \cdots, w-1$. However, the solutions with $l > (w-1)/2$ are redundant, because the solutions for $l$ and $l'$ with $l + l' = w - 1$ are actually the same one, as can been appreciated from Eq.(7). Therefore, the total number of solutions with $\psi_4 = \psi_1^*$ is the number of nonnegative integer smaller than $(w-1)/2$, which is $[w/2]$

(Here, "$[\cdots]$" denotes the floor function, mapping a real number to the largest previous integer), excluding possible solution with $l = (w-1)/2$.

Now we consider the other choice: $\psi_4 = -\psi_1^*$. By calculations similar to the case $\psi_4 = \psi_1^*$, we can obtain equations similar to Eq.(5), except that the "+" sign before $m_0$ is replaced by "−". We adopt the same ansatz as given in Eq.(7), and follows the steps below Eq.(7), solving the case $l \neq (w-1)/2$ first. It is found that the number of solutions is $[w/2]$.

Finally, we study the special case $l = (w-1)/2$ (This case needs consideration only when $w$ is odd; for $w$ even, this option is automatically absent). Given this value of $l$, the second term in Eq.(7) becomes redundant, thus we can take

$$\psi_1^{(l)} = e^{i\pi/4} u_l e^{il\theta}. \tag{10}$$

Under the choice $\psi_4 = \pm\psi_1^*$, we obtain the single differential equation

$$v(\frac{\partial}{\partial r} - \frac{l}{r})u_l \pm m_0 u_l = 0. \tag{11}$$

For the choice "+" of the "$\pm$", Eq.(11) has a single normalizable solution with asymptotic behaviors $u_l \to r^l$ in the $r \to 0$ limit and $u_l \to \exp(-\frac{m_0}{v}r)$ in the $r \to \infty$ limit. For the choice "−" of the "$\pm$", Eq.(11) leads to $u_l \to \exp(\frac{m_0}{v}r)$ in the $r \to \infty$ limit, which is apparently not normalizable. Therefore, there exists a single normalizable localized mode, in the $\psi_4 = \psi_1^*$ sector, for the special case $l = (w-1)/2$. We also note that, if we take $m_0 < 0$ instead of $m_0 > 0$, the normalizable solution would be present in the $\psi_4 = -\psi_1^*$ sector but absent in the $\psi_4 = \psi_1^*$ sector, thus the total number of solution is the same.

Let us summarize the above calculations as follows. When $w$ is odd, the total number of normalizable solutions is $2[w/2]+1 = w$; when $w$ is even, the total number of normalizable solutions is $2[w/2] = w$. Therefore, *the total number of topological modes is always $w$, irrespective of the parity (odd/even) of $w$*. Furthermore, our calculation shows that the eigenvalues take the simple form

$$E(k_z) = vk_z, \tag{12}$$

thus all these $w$ modes propagate along the $+z$ direction, with the same velocity $v$.

Finally, let us discuss the one-way modes for the integer $w < 0$. Solution of the form of $\psi = [\psi_1, 0, 0, \psi_4]^T$ does not exist in this case, because the condition given in Eq.(9) can never be satisfied. On the other hand, solutions of the form of $\psi = [0, \psi_2, \psi_3, 0]^T$ can be found. In fact, we can follow the steps above and obtain the equations

$$\begin{aligned}
-ive^{-i\theta}(\frac{\partial}{\partial r} - \frac{i}{r}\frac{\partial}{\partial \theta})\psi_2 + m_0 e^{iw\theta}\psi_3 &= 0, \\
m_0 e^{-iw\theta}\psi_2 + ive^{i\theta}(\frac{\partial}{\partial r} + \frac{i}{r}\frac{\partial}{\partial \theta})\psi_3 &= 0,
\end{aligned} \tag{13}$$

whose complex conjugations are

$$\begin{aligned}
-ive^{i\theta}(\frac{\partial}{\partial r} + \frac{i}{r}\frac{\partial}{\partial \theta})\psi_2 + m_0 e^{i|w|\theta}(-\psi_3) &= 0, \\
m_0 e^{-i|w|\theta}\psi_2 + ive^{-i\theta}(\frac{\partial}{\partial r} - \frac{i}{r}\frac{\partial}{\partial \theta})(-\psi_3) &= 0.
\end{aligned} \tag{14}$$

We can see that Eqs.(14) are the same as Eqs.(4) except that $\psi_2$ and $-\psi_3$ take the place of $\psi_1$ and $\psi_4$, respectively. Now our previous analysis for Eqs.(4) with $w \geq 1$ immediately tells us that the number of one-way modes for $w < 0$ is $|w|$. Because the solutions for $w < 0$ take the form of $\psi = [0, \psi_2, \psi_3, 0]^T$, the dispersion is $E(k_z) = -vk_z$, thus all the one-way modes propagate in the $-z$ direction.

## Calculation of the second Chern number $C_2$

The effective Hamiltonian Eq.(2) takes the form of

$$H_{\text{eff}} = \sum_{a=1}^{5} d_a \Gamma^a, \tag{15}$$

where the Dirac matrices $\Gamma^a = \sigma_a \tau_z$ ($a = 1, 2, 3$), $\Gamma^4 = \tau_x$, $\Gamma^5 = \tau_y$, the coefficient functions $d_a = v k_a$ ($a = 1, 2, 3$), $d_4 = \text{Re}(m)$, $d_5 = -\text{Im}(m)$. A straightforward calculation[4] of $C_2$ leads to

$$\begin{aligned} C_2 &= \frac{3}{8\pi^2} \int d\theta d^3k \, \epsilon^{abcde} \hat{d}_a \frac{\partial \hat{d}_b}{\partial k_x} \frac{\partial \hat{d}_c}{\partial k_y} \frac{\partial \hat{d}_d}{\partial k_z} \frac{\partial \hat{d}_e}{\partial \theta} \\ &= \frac{1}{2\pi} \int_0^{2\pi} d\theta \frac{d[\arg(m(\theta))]}{d\theta}, \end{aligned} \quad (16)$$

where $\hat{d}_a = d_a / \sqrt{\sum_{b=1}^5 d_b^2}$. With the Dirac mass $m = m_0 \exp(iw\theta)$, we have

$$C_2 = w. \quad (17)$$

For a general modulation that combines several different spatial frequency, namely, $m(\theta) = \sum_w m_w \exp(iw\theta)$, Eq.(16) is not amenable to further simplification in the generic cases, however, we have $C = w_0$ in the cases that $|m_{w_0}| > \sum_{w \neq w_0} |m_w|$, in other words, $C_2$ is determined by the dominant modulation.

---

* Electronic address: linglu@iphy.ac.cn
† Electronic address: wangzhongemail@tsinghua.edu.cn



\end{document}